\documentclass[15]{article}
\usepackage{amsmath}
\usepackage{graphics}
\usepackage{graphicx}

\newcommand{\NumOpWithLow}[2]{\hat{#1}^\dag_#2\hat{#1}_#2}
\newcommand{\NumSum}[1]{\sum_{j=x,y,z}\NumOpWithLow{#1}{j}}
\newcommand{\DefNumSum}{\NumSum{a}}

\newcommand{\Ket}[1]{\left\vert#1\right\rangle}

\newcommand{\BraKet}[2]{\left\langle#1\right\vert\left.#2\right\rangle}
\newcommand{\KetBra}[2]{\left\vert#1\right\rangle\left\langle#2\right\vert}
\newcommand{\MatrixEl}[3]{\left\langle#1\right\vert\hat{#2}\vert\left.#3\right\rangle}
\newcommand{\UnitarEl}[3]{\left\langle#1\right\vert {#2}\vert\left.#3\right\rangle}

\newcommand{\vett}[1]{\stackrel{\rightarrow}{#1}}

\newcommand{\HZero}[3]{\hbar #1\DefNumSum+\sum_{i=1,#3}\hbar #2_i\KetBra{i}{i}}
\newcommand{\HInt}[4]{#1\vett{a}^{\vett{#2}}\KetBra{2}{1}+#3\vett{a}^{\vett{#4}}\KetBra{3}{2}}

\begin{document}

\title{Quantum Zeno Effect in Trapped Ions}
\author{B.Militello, A.Messina and A.Napoli\\ \emph{INFM and MURST}\\ \emph{Dipartimento di Scienze Fisiche ed                         Astronomiche,}\\ \emph{via Archirafi 36, 90123 Palermo, Italy}}
\date{\today}
\maketitle{}

\section*{Abstract}

\small{A "continuous measurement" Quantum Zeno Effect (QZE) in the context of trapped ions is predicted. We describe the physical system and study its exact time evolution showing the appearance of Zeno Phenomena. New indicators for the occurrence of QZE in oscillatory systems are proposed and carefully discussed.\\
PACS : 03.65.Bz; 32.80.Pj; 42.50.Ct\\
KeyWords : Quantum Zeno Effect, Trapped Ion, Three-level
systems\\}


Over the last few years the quantum dynamics of many models describing trapped ions have been investigated. The interest on this subject follows from the fact that such physical systems permit the observation of noteworthy quantum phenomena like, for example, quantum interference in Schrodinger cat states[1], decoherence[2,3,4] and quantum chaos[5,6]. In this letter our attention is focused on the quantum Zeno effect (QZE). In its standard form it consists of an inhibition of the system dynamics due to a continuous measurement process[7]. Misra and Sudarshan have in fact proved that in the limit of an infinite number of measurements performed in a fixed time interval, the system is frozen in its initial state[8]. Sometimes it happens that an appropriately large but finite number of measurements makes the initial state abandonment easier. In these cases we speak about Inverse Zeno Effect (IZE)[9,10].\\
Quite recently Facchi and Pascazio [9] have introduced a broader definition of Zeno effect which in this paper we refer to as Generalized QZE (GQZE). They consider the most general case in which an additional interaction, that may even not be a measurement interaction, is responsible for hindering the system dynamics.
In order to better understand the physical meaning of GQZE, let us recall its definition more in detail.
Suppose a physical system under scrutiny describable in terms of a hamiltonian model\\
\begin{eqnarray}
H_K = H + H_{meas}(K)
\end{eqnarray}
where $K$ is a parameter and $H_{meas}(K)$ is an interaction term which may even be associated with a measurement process and such that $H_{meas}(K=0)=0$.
By definition a GQZE occurs if, for $K\not=0$, there exists a time interval $[t_{1}^K,t_{2}^K]$ in which the survival probability is greater than the one correspondent to $K=0$:
\begin{eqnarray}
P^{K}(t) > P(t)
\end{eqnarray}
where
\begin{eqnarray}
P^{K}(t) = |\BraKet{\psi(0)}{\psi(t)}|^2 = |\UnitarEl{\psi (0)}{\hat U_K}{\psi (0)}|^2
\end{eqnarray}
($\hat U_K$ being the time evolution operator relative to $H_K$) is the probability to find the system in its initial state $\Ket{\Psi(0)}$, and
\begin{eqnarray}
P(t) \equiv P^{K=0}(t)
\end{eqnarray}
It is worth noting that this definition is applicable to both unstable and oscillatory systems. In these physical situations one can single out a Poincar\'e time, $T_p(K)$, finite in the latter case and infinite in the former one. To claim the occurrence of a GQZE in oscillatory systems, in accordance with ref[9], one must require that $|t_{1}^K-t_{2}^K|$ is of the same order of $T_p(K)$. \\
In this paper we show the occurrence of a GQZE in the dynamics of a laser driven trapped ion. A formally equivalent result was found by Peres[11] in a different physical situation, while a standard QZE was brought to the light theoretically by Cook    et al. and, experimentally, by Wineland et al.[12] in the context of trapped ions.\\
Novel ways to characterize the appearance of Zeno Phenomena in oscillatory systems are moreover introduced and discussed in detail.\\
Our physical system consists of a three-level ion confined in an isotropic three dimensional Paul trap interacting with a set of laser fields appropriately directed, polarized and tuned. \\
In a Paul trap an inhomogeneous time dependent electric field  and a quadrupolar static electric field confine a charged particle in a finite region of space wherein the ion moves subjected to an effectively three dimensional harmonic isotropic potential of frequency $\omega_0$ [13,14].\\
The hamiltonian model describing such a situation is
\begin{eqnarray}
  H_0 = \HZero{\omega_0}{\omega}{3}
\end{eqnarray}
where the annihilation (creation) operator $\hat{a}_j({\hat{a}_j}^\dag)$ describes the harmonic motion of the center of mass of the trapped ion along the axes $j$ ($j=x,y,z$) and the $i$-th internal level and its energy are denoted by $\Ket{i}$ and $\hbar\omega_i$ respectively with $i$=1,2,3.\\
A transition frequency $\omega$ between energy levels of $H_0$, involving the passage of the atom from the internal state $\Ket{i}$ to the internal one $\Ket{j}$ together with a simultaneously motional state modification is called $m-$th red(blue) sideband when $\omega-(\omega_j-\omega_i)\equiv\omega-\omega_{ij}=m\omega_0$ is negative(positive). If $m=0$, that is when no bosonic transition is involved, $\omega$ is called carrier frequency.\\
Let us consider now the application on the trapped ion of a laser tuned to the first red sideband of the atomic transition 1-2 and another laser tuned to the first red sideband of the transition 2-3. The first laser is chosen of frequency $\omega_a=\omega_{12}-\omega_0$, wave vector $k_a$, initial phase $\phi_a$ and directed along $x$ and the second one, directed along $y$, characterized by $\omega_b=\omega_{23}-\omega_0$, $k_b$, $\phi_b$.
The presence of the two lasers causes the appearance of a coupling between internal (electronic state) and external (center of mass) degrees of freedom of the trapped ion[15].\\
Supposing $\omega_a,\omega_b\gg\omega_0$, then the most important interaction term is the dipole one which in the Schrodinger picture can be cast in the form
\begin{eqnarray}
  \nonumber H^s_1 &=& -\vett{d}\cdot\vett{E} = \\&=&(\sum_{i,j=1,3}\vett{d}_{ij}\KetBra{i}{j})(\sum_{l=a,b}\vett{E}_l e^{i(k_l \hat{x} - \omega_l t   + \phi_l)}+(h.c.))
\end{eqnarray}
where $\vett{E}_a$ and $\vett{E}_b$ describe amplitudes and polarizations of the applied fields, $\vett{d}$ is the atomic dipole operator and $\vett{d}_{ij}$ is the matrix element of $\vett{d}$ between $\Ket{i}$ and $\Ket{j}$.\\
In the interaction picture relative to the unperturbed hamiltonian $H_0$, expanding all the exponential operators and recalling that $\hat{x}(\hat{y})=\epsilon(\hat{a}_{x(y)}+\hat{a}_{x(y)}^{\dag})$, we get
\begin{eqnarray}
  \nonumber H^I_1 = (\hbar\Omega_a e^{-\frac{\eta^2_a}{2}}e^{i\phi_a}\sum_{j=0,\infty}\frac{(i\eta_a)^{2j+1}}{j!(j+1)!}     \hat{a}_x^j(\hat{a}_x^\dag)^{j+1}\sigma_{12} +\\
  \hbar\Omega_b e^{-  \frac{\eta^2_b}{2}}e^{i\phi_b}\sum_{j=0,\infty}\frac{(i\eta_b)^{2j+1}}{j!(j+1)!}\hat{a}_y^j(\hat{a}_y^\dag)^{j+1}      \sigma_{23} + hc)
\end{eqnarray}
after having discarded all the anti-rotating terms (Rotating Wave Approximation).\\
In eq.(7) $\Omega_a=\frac{\vett{E}_a\cdot\vett{d}_{12}}{\hbar}$, $\Omega_b=\frac{\vett{E}_b\cdot\vett{d}_{23}}{\hbar}$, $\eta_l=k_l\epsilon$ with $l=a,b$ and $\epsilon=\sqrt{\frac{\hbar}{2M\omega_0}}$, $M$ being the mass of the ion.
The quantities  $\eta_a,\eta_b$ measure the order of magnitude of the center of mass oscillation amplitudes with respect to the fields wavelengths.
Finally $\sigma_{ij}\equiv\KetBra{i}{j}$ is the atomic transition operator.\\
If $\eta_a,\eta_b\ll1$ (Lamb-Dicke approximation), we are legitimated to truncate the series above up to the terms linear in $\eta_a$ and $\eta_b$, obtaining
\begin{eqnarray}
  H^I_1 = \gamma_1 {a_x}\KetBra{2}{1}+\gamma_2 {a_y}\KetBra{3}{2}+(h.c.)
\end{eqnarray}
where the initial phases $\phi_a$ and $\phi_b$ are chosen equal to $\frac{\pi}{2}$ to make real
the quantities $\gamma_1=-\hbar\Omega_a e^{-\frac{\eta^2_a}{2}}\eta_a$
and $\gamma_2=-\hbar\Omega_b e^{-\frac{\eta^2_b}{2}}\eta_b$.\\
Implementing this hamiltonian model is certainly in the grasp of experimentalists. Generally speaking, one of the most attractive aspects of trapped ions systems is the relative easiness with which specific models describing the interaction between few level systems and few bosons may be realized, with respect to a lot of other physical situations like, for example, CQED[15].\\
In eq.(6) we have considered the sum of two laser fields and this leads to the effective hamiltonian model given by eq.(8). Increasing the number of laser fields and different choosing their frequencies, wavevectors, initial phases and amplitudes, we may realize effective hamiltonian models more and more complicated. For example, we may consider the physical situation wherein, in place of the laser 1-2 red sideband tuned previously used, we take two lasers tuned to the second 1-2 red sideband, $\pi$ out of phase, with the same amplitudes and directed along the axis $x'$ and $y'$ rotated of $\frac{\pi}{4}$ with respect to the axis $x$ and $y$. The hamiltonian model describing such a physical system may be written down as $H^I_1 = \gamma_1 {a_x a_y}\KetBra{2}{1}+\gamma_2 {a_y}\KetBra{3}{2}+(h.c.)$ [16].\\
On the basis of this example it is not difficult to convince oneself that it is always possible to engineer an appropriate laser beam configuration such that the vibronic coupling in the interaction picture may be represented in terms of the following general model
\begin{eqnarray}
  H_1 = \HInt{\gamma_1}{r}{\gamma_2}{l}+(h.c.)
\end{eqnarray}
where
\begin{eqnarray}
\vett{a}^{\vett{p}}\equiv a_x^{p_x}a_y^{p_y}a_z^{p_z}\;\;\;\;\;\;\;\;(\vett{p}=\vett{r},\vett{l})
\end{eqnarray}
and $\vett{r},\vett{l}$ are sets of three natural numbers.\\
This is the hamiltonian model, on which we focus from now on. It generalizes eq.(8) in the sense that more bosons and all the three vibrational modes are involved in the process. The related dynamical problem may be exactly solved for arbitrary $\vett{r}$ and $\vett{l}$.\\
Let us indicate the eigenstates of $H_0$ with

\begin{eqnarray}
\Ket{\vett{n},\sigma}=\Ket{n_x,n_y,n_z,\sigma}
\end{eqnarray}
where $\sigma=1,2,3$ refers to the atomic states.\\
It is possible to see that in this basis the operator $H_1$ is diagonal in blocks which are one dimensional if $n_i-r_i < 0$, two dimensional if $n_i-r_i \geq 0 \;\;\; and  \;\;\; n_i-r_i-l_i < 0$, and three dimensional if $n_i-r_i-l_i \geq 0$ with $i=x,y,z$.\\
Here our attention is centered on a 3x3 representation of $H_1$ having the form

\begin{eqnarray}
\left(
\begin{array}{ccc}
0&\alpha&0\\\
\alpha^*&0&\beta\\
0&\beta^*&0
\end{array}
\right)
\end{eqnarray}
in the subspace generated by $\Ket{\vett{n},1}$, $\Ket{\vett{n}-\vett{r},2}$, $\Ket{\vett{n}-\vett{r}-\vett{l},3}$.\\
The non vanishing matrix elements appearing above are defined as follows

\begin{eqnarray}
\alpha\equiv\MatrixEl{\vett{n},1}{H_1}{\vett{n}-\vett{r},2} = \gamma_1\;
\sqrt{\prod_{i=x,y,z}\frac{n_i !}{(n_i-r_i)!}}
\end{eqnarray}

\begin{eqnarray}
\beta\equiv\MatrixEl{\vett{n}-\vett{r},2}{H_1}{\vett{n}-\vett{r}-\vett{l},3} = \gamma_2\;
\sqrt{\prod_{i=x,y,z}\frac{(n_i-r_i)!}{(n_i-r_i-l_i)!}}
\end{eqnarray}

The parameters $\alpha$ and $\beta$ measure the effective strength of the couplings between the three levels and are $\vett{n}$-,$\vett{r}$- and $\vett{l}$-dependent.\\
If $r_i=l_i=0$ for $i=x,y,z$, then the dynamical system described by the matrix (12), involves only the three electronic levels of the trapped ion. When, on the contrary, at least one of the laser is tuned to a sideband and the initial state of the center of mass is a Fock state such that $n_i-r_i-l_i\ge 0$ for all $i=x,y,z$, then the hamiltonian matrix (12) describes the dynamics of a truly vibronic three-level system.\\
It is straightforward to get the following time evolutions (in the interaction picture):

\begin{eqnarray}
\nonumber\Ket{\vett{n},1}\rightarrow\psi_{\vett{n},1}(t)&=&\frac{|\beta|^2+|\alpha|^2\cos{\omega t}}{\tilde{\epsilon}^2}\Ket{\vett{n},1}-i\frac{\alpha^*}{\tilde{\epsilon}}\sin{\omega t}\Ket{\vett{n}-\vett{r},2}+\\
&+&\frac{\alpha^*\beta^*}{\tilde{\epsilon}^2}(\cos{\omega t}-1)\Ket{\vett{n}-\vett{r}-\vett{l},3}
\end{eqnarray}
\begin{eqnarray}
\nonumber\Ket{\vett{n}-\vett{r},2}\rightarrow\psi_{\vett{n}-\vett{r},2}(t)&=&\cos{\omega t}\Ket{\vett{n}-\vett{r},2}-i\frac{\alpha}{\tilde{\epsilon}}\sin{\omega t}\Ket{\vett{n},1}+\\
&-&i\frac{\beta^*}{\tilde{\epsilon}}\sin{\omega t}\Ket{\vett{n}-\vett{r}-\vett{l},3}
\end{eqnarray}
\begin{eqnarray}
\nonumber\Ket{\vett{n}-\vett{r}-\vett{l},3}\rightarrow\psi_{\vett{n}-\vett{r}-\vett{l},3}(t)=
\nonumber\frac{\alpha\beta}{\tilde{\epsilon}^2}(\cos{\omega t}-1)\Ket{\vett{n},1}+\\
-i\frac{\beta}{\tilde{\epsilon}}\sin{\omega t}\Ket{\vett{n}-\vett{r},2}+
\frac{|\alpha|^2+|\beta|^2\cos{\omega t}}{\tilde{\epsilon}^2}\Ket{\vett{n}-\vett{r}-\vett{l},3}
\end{eqnarray}
where
\begin{eqnarray}
\tilde{\epsilon}=\sqrt{|\alpha|^2+|\beta|^2}=|\alpha|\sqrt{1+|\chi|^2}
\end{eqnarray}
\begin{eqnarray}
\omega=\omega(\chi)=\frac{\epsilon}{\hbar}=\frac{|\alpha|}{\hbar}\sqrt{1+|\chi|^2}
\end{eqnarray}
with
\begin{eqnarray}
\chi = \frac{\beta}{\alpha} = \frac{\gamma_2}{\gamma_1}\prod_{i=x,y,z}\frac{n_i-r_i!}{\sqrt{n_i!(n_i-r_i-l_i)!}}
\end{eqnarray}
The quantity $|\chi|$ expresses the relative strength of the coupling between the level $\Ket{\vett{n}-\vett{r},2}$ with  $\Ket{\vett{n}-\vett{r}-\vett{l},3}$ and $\Ket{\vett{n},1}$ respectively and hence, as $\alpha$ and $\beta$, is $\vett{n}$-,$\vett{r}$- and $\vett{l}$-dependent and is determined both by laser parameters and by the initial condition.\\
Such a property enables to obtain larger values of $|\chi|$ with the same laser amplitudes only choosing different initial vibrational states, and hence allows to avoid possible difficulties related with the use of very intense laser fields.\\
In passing we note that starting from the initial 3D zero point energy vibrational state, the Boulder group has succeeded in generating one-dimensional Fock states up to 16 bosonic excitations.[15]\\
In the limit $|\chi| \gg 1$, the time evolution of the state $\Ket{\vett{n},1}$, as given by eq.(15), appears to be inhibited. This suggests us that when the coupling between levels 2 and 3 is greater than the one between 1 and 2, a GQZE occurs.\\
In order to demonstrate the occurrence of such an effect in our system, in accordance with ref[9], let us compare our hamiltonian model given by eq.(9) with that given by eq.(1) making the correspondences
$H\rightarrow\gamma_1 {\vett{a}}^{\vett{k}} \KetBra{2}{1}+(h.c.)$, $\chi\propto K\rightarrow\frac{\gamma_2}{\gamma_1}$ and $H_{meas}(K)\rightarrow K \gamma_1 {\vett{a}}^{\vett{l}} \KetBra{3}{2}+(h.c.)$.\\
Consider now the survival probability, $P^{\chi}(t)$, for different values of $\chi$, with $\Ket{\vett{n},1}$ as initial state. Since the initial state is an eigenstate of $H_0$ we have
\begin{eqnarray}
P^{\chi}(t) \equiv |\UnitarEl{\vett{n},1}{e^{\frac{i}{\hbar}H_0t}}{\psi_{\vett{n},1}}|^2 =|\BraKet{\vett{n},1}{\psi_{\vett{n},1}}|^2
\end{eqnarray}
and therefore
\begin{eqnarray}
P^{\chi}(t)=\left(\frac{|\chi|^2+\cos{\omega t}}{|\chi|^2+1}\right)^2
\end{eqnarray}
In view of eqs.(15)-(17) the time evolution of the trapped ion within the subspace under scrutiny, is governed by a single frequency $\omega(\chi)$. This naturally leads to define the Poincar\'e-time, $T_p(\chi)$, of our system as the oscillation period, $T(\chi)$,
\begin{eqnarray}
T_p(\chi)=T(\chi)=\frac{2\pi}{\omega(\chi)}=\frac{2\pi\hbar}{|\alpha|\sqrt{1+|\chi|^2}}
\end{eqnarray}
Thus $P^{\chi}(0)=P^{\chi}(T_p(\chi))=1$, which simply means that after a complete $\chi$-dependent Rabi cycle the system is found in its initial state.\\
Figure 1 shows that for each value of $|\chi|>1$ there exists a time instant $t_{\chi}$ of the order of $T_p(\chi)$ such that for $t\epsilon[0,t_{\chi}]$ we have $P^{\chi}(t) > P(t)$. Thus,  at the light of ref[9], we may claim that our system manifests GQZE, under the condition previously discussed.\\
It is worth noting that $P^{\chi}(t)$ exhibits an absolute minimum, $m(\chi)$, in the time interval $[0,T_p(\chi)]$, which may easily be evaluated as
\begin{equation}
m(\chi) =
  \begin{cases}
    0 &  \text{if $0\le|\chi|\le 1$},\\
    \left(\frac{|\chi|^2-1}{|\chi|^2+1}\right)^2 &  \text{if $|\chi|> 1$},\\
  \end{cases}
\end{equation}
and occurs, for the first time, at
\begin{equation}
t_m(\chi) =
  \begin{cases}
    \frac{\hbar}{|\alpha|\sqrt{1+|\chi|^2}}arccos(-|\chi|^2) &  \text{if $0\le|\chi|\le 1$},\\
    \frac{\pi\hbar}{|\alpha|\sqrt{1+|\chi|^2}} &  \text{if $|\chi|> 1$},\\
  \end{cases}
\end{equation}
In figures 2 and 3 we plot the behaviours of $m(\chi)$ and $t_m(\chi)$ respectively.\\
The value of $|\chi|$ at which $m(\chi)$ begins to assume values different from zero is $|\chi|=1$ and corresponds to the position of the maximum exhibited by $t_m(\chi)$. It is interesting to note, in passing, that the non differentiability of $t_m(\chi)$ at $|\chi|=1$ originates from the coalescence of three stationary points (two minima and a maximum) of $P^{\chi}(t)$[17].\\
We wish to emphasize that, as $|\chi|$ grows up, the value of $m(\chi)$ tends toward 1 and hence $P^{\chi}(t)$ tends toward 1 uniformly with respect to $\chi$. Such a property, namely $m(\chi)\rightarrow 1$ as $|\chi|\rightarrow\infty$, describes in a very transparent way the effectiveness of the hindering process of the initial state time evolution as a function of $\chi$, and provides a new way to bring to the light the occurrence of a GQZE in physical systems characterized by a finite Poincar\'e-time.\\
It should be noted that this definition is stronger than the one given by Facchi and Pascazio[9]. This virtue proceeds from the simplicity of our physical system, which permits us to analyze in detail its dynamical properties and from the fact that we have restricted our GQZE-testing to oscillatory phenomena.\\
Another way to construct an indicator of the GQZE-occurrence in finite Poincar\'e-time systems is now proposed. Instead of considering the minimum of $P^{\chi}(t)$, we may use the mean value of the survival probability in a $\chi$-dependent Rabi cycle :
\begin{eqnarray}
\overline{P}(\chi) = \int^{T(\chi)}_0{P^{\chi}(t)\frac{dt}{T(\chi)}}= \frac{\chi^4+\frac{1}{2}}{(1+\chi^2)^2}
\end{eqnarray}\\
This quantity has a very clear physical meaning representing the probability to find the system in its initial state independently of the time instant randomly chosen in $[0,T(\chi)]$. \\
The behaviour of $\overline{P}(\chi)$ is plotted in fig.4 and shows the existence of a minimum at $|\chi|=\frac{1}{\sqrt{2}}$ and the asymptotic tendency to 1 when $|\chi|$ becomes larger and larger. This asymptotic behaviour may be reasonably taken as a clear indication that the system dynamics becomes more and more hindered "in average".
In order to  elucidate this statement, it is useful to introduce the quantity $S(\chi,\epsilon)$ defined as the sum of amplitudes of the time intervals included in $[0,T_p(\chi)]$ wherein $P^{\chi}(t) < \overline{P}(\chi) - \epsilon$, $\epsilon$ being a positive arbitrarily chosen $\chi$-independent real quantity.\\
By "average hindering" we mean the tendency of the ratio $\frac{S(\chi,\epsilon)}{T_p(\chi)}$ to vanish as $\overline{P}(\chi)\rightarrow 1$.\\
Intuitively, since each $P^{\chi}(t)$ is upper-bounded by 1, if its temporal mean value, $\overline{P}(\chi)$, approaches this maximum, the function $P^{\chi}(t)$ tends to 1 almost anywhere.\\
It is interesting to note that introducing, in analogy with
eq.(26), the temporal mean value of the probability
$\overline{P_2}(\chi)$ ($\overline{P_3}(\chi)$) of finding the
trapped ion in the level $\Ket{\vett{n}-\vett{r},2}$
($\Ket{\vett{n}-\vett{r}-\vett{l},3}$), the position of the
absolute minimum of $\overline{P}(\chi)$ is just that in
correspondence to which
\begin{eqnarray}
\overline{P}(\chi) = \overline{P_2}(\chi) = \overline{P_3}(\chi) = \frac{1}{3}
\end{eqnarray}
where the last equality follows from the normalization condition $\overline{P}(\chi)+\overline{P_2}(\chi)+\overline{P_3}(\chi)=1$.\\
We wish to point out that the $\overline{P}(\chi)$-test is weaker than the $m(\chi)$-test, in the sense that when $m(\chi)$ grows up and approaches to 1, the parameter $\overline{P}(\chi)$ approaches to 1 too but not vice versa. Nevertheless introducing $\overline{P}(\chi)$ appears to be useful in order to bring to the light a "global decrease" of $P^{\chi}(t)$ which is invisible when $m(\chi)$-test is considered.\\
Summarizing, in this paper we have predicted that an ion confined in a Paul trap may exhibit a QZE when it is irradiated with appropriately configured laser beams. An important point that deserves to be emphasized is that our prediction concerns a continuous measurements QZE which is different from the QZE "\'a la Misra-Sudarshan" predicted by Cook and experimentally found by Wineland[12]. The difference is that in our scheme only unitary evolutions are involved while in Cook's experiment description projection operators are used and also the interaction with a series of short separated pulses is required. But the shorter is the distance between two pulses the more unfeasible is the series. On the contrary, a continuous interaction is simpler to realize and hence our scheme might be relatively easier to implement.\\
The simplicity of the dynamical problem considered in this paper has given us the possibility of proposing novel indicators for the occurrence of the QZE. We feel that such novel ways of analysing the presence of QZE might be useful for different physical contexts provided that their dynamics is characterized by a finite recurrence time.


\section*{Acknowledgements}

We wish to thank Paolo Facchi and Saverio Pascazio for reading the manuscript and for giving us many useful suggestions.
The financial support from CRRNSM-Regione Sicilia is greatly acknowledged. One of the author (A.N.) is indebted to MURST and FSE for partially supporting this in the context of Programma Operativo Multiregionale 1994/99 94002311.


\section*{References}

$[1]$  C.Monroe, D.J.Wineland et al., Science 272(1996) 1131\\
$[2]$  S.Schneider and G.J.Milburn, Phys. Rev. A 57 (1998) 3748\\
$[3]$  S.Schneider and G.J.Milburn, Phys. Rev. A 59 (1999) 3766\\
$[4]$  Le-Man Kuang et al., Phys. Rev. A 60 (1999)3815\\
$[5]$  S.A.Gardiner, J.I.Cirac and P.Zoller, Phys. Rev. Lett. 79 (1997) 4790\\
$[6]$  A.J.Scott, C.A.Holmes, G.J.Milburn Phys. Rev. A 61 (1999) 13401\\
$[7]$  K. Machida, H.Nakazato, S. Pascazio et al, Phys. Rev. A 60 (1999) 3448\\
$[8]$  B. Misra and C.G.Sudarshan, J. Math. Phys. 18 (1977)756 \\
$[9]$  P.Facchi and S. Pascazio, Quantum Zeno and Inverse Quantum Zeno Effects, in \emph{Progress in Optics}, 42 (2001) ed. E.Wolf, Elsevier, Amsterdam\\
$[10]$ P.Facchi and S. Pascazio, Phys. Rev. A 62 (2000) 23804\\
$[11]$ A.Peres, Am. J. Phys. 48 (1980) 931\\
$[12]$ Wineland et al., Quantum Zeno Effect, Phys. Rev. A 41 (1990) 2295\\
$[13]$ P.K.Ghosh, Ion Traps, Clarendon Press, Oxford 1995\\
$[14]$ P.E.Toschek, in \emph{New trends in atomic physics} ed G.Grynberg and R. Stora, Elsevier Science Publishers B.V. (1984) 383\\
$[15]$ D.J.Wineland, C. Monroe, W.M.Itano, et al., J. Res. N.I.S.T. 103(1998) 259\\
$[16]$ S.Maniscalco, A.Messina and A.Napoli, Phys. Rev. A 61 (2000) 53806\\
$[17]$ We thank Paolo Facchi for suggesting us this interpretation
and its connection with the catastrophe theory presented for
example in V.I.Arnold, Catastrophe Theory, Springer-Verlag, Berlin
1984
\pagebreak

\begin{figure}[h!]
\centering
\includegraphics[width=10cm]{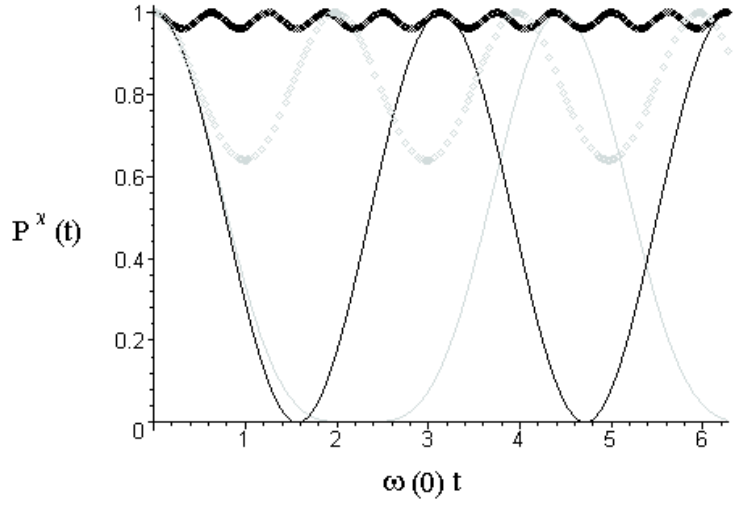}
\caption{Survival probability $P^{\chi}(t)$ vs $\frac{1}{\omega(0)}$-scaled time for different values of $|\chi|$. $|\chi|=0$ (black line), $|\chi|=1$ (grey line), $|\chi|=5$ (grey dots), $|\chi|=10$ (black dots)}
\end{figure}
\pagebreak

\begin{figure}[h!]
\centering
\includegraphics[width=10cm]{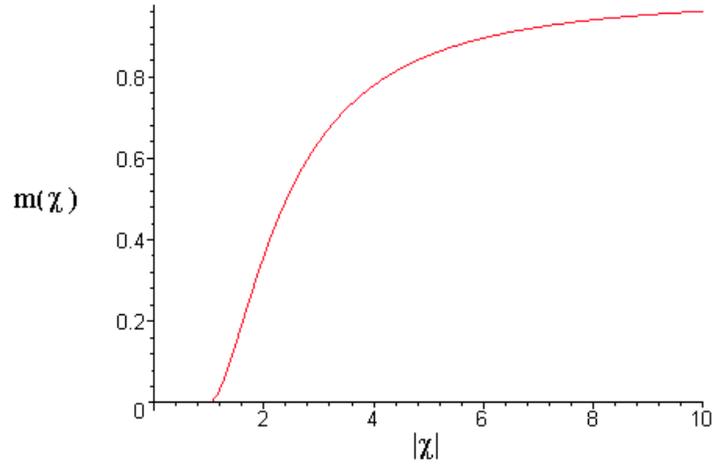}
\caption{Minimum $m(\chi)$ of the survival probability vs $|\chi|$}
\end{figure}
\pagebreak

\begin{figure}[h!]
\centering
\includegraphics[width=10cm]{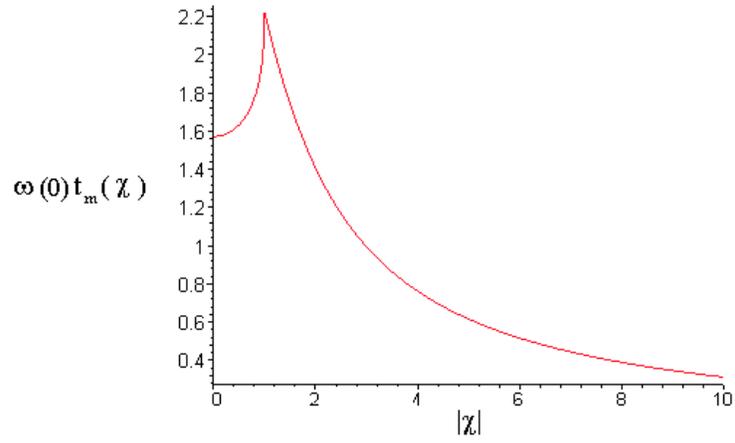}
\caption{Time $t_m(\chi)$ ($\frac{1}{\omega(0)}$-scaled) at which the first minimum $m(\chi)$ of the survival probability $P^{\chi}(t)$ occurs vs $|\chi|$}
\end{figure}
\pagebreak

\begin{figure}[h!]
\centering
\includegraphics[width=10cm]{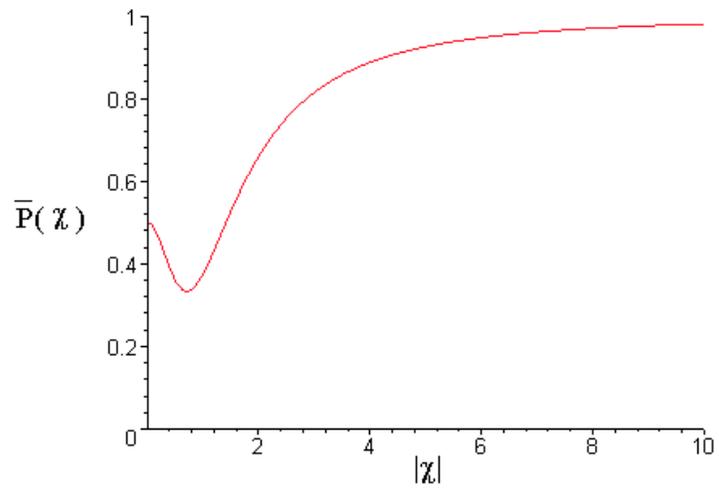}
\caption{Temporal mean value of the survival probability, $\overline{P}(\chi)$, vs $|\chi|$}
\end{figure}

\end{document}